\documentclass[%
 reprint,
 amsmath,amssymb,
 aps,
]{revtex4-2}

\usepackage{graphicx}
\usepackage{dcolumn}
\usepackage{bm}

\begin{document}

\preprint{APS/123-QED}

\title{Transition from non-ergodic to ergodic dynamics in an autonomous discrete time crystal
}

\author{T. T. Sergeev}
\author{A. A. Zyablovsky}
\email{zyablovskiy@mail.ru}
\author{E. S. Andrianov}\

\affiliation{Dukhov Research Institute of Automatics, 127055, 22 Sushchevskaya, Moscow, Russia}
\affiliation{Moscow Institute of Physics and Technology, 141700, 9 Institutskiy pereulok, Moscow, Russia}
\affiliation{Institute for Theoretical and Applied Electromagnetics, 125412, 13 Izhorskaya, Moscow, Russia}

\date{\today}

\begin{abstract}
  We consider an autonomous system of two coupled single-mode cavities, one of which interacts with a multimode resonator. We demonstrate that for small coupling strengths between single-mode cavities, the Loschmidt echo oscillates periodically in time and spontaneous breaking of time translation symmetry takes place. The Loschmidt echo behavior is an indication of the non-ergodic nature of the system when its evolution is time-reversible and the system retains a memory of the initial state under the action of small perturbations. This behavior reveals the presence of a time crystalline order in the autonomous system. In this regime, the system is a new class of time crystals - autonomous discrete time crystals. An increase in the coupling strength leads to a transition from periodic oscillations to an exponential decay in time of the Loschmidt echo. This corresponds to the transition from non-ergodic behavior to ergodic one in the system, and is accompanied by the disappearance of time crystalline order. We demonstrate that at the transition point the time-averaged variance of the number of photons reaches a maximum, which serves as a signature of the transition. We show that such a transition can also be observed when changing the number of degrees of freedom in the resonator, which is achieved by changing its length.
   
\end{abstract}

\maketitle

\section*{Introduction}
The time evolution of closed quantum systems is determined by the Schrödinger equation, which is time-reversible \cite{31}. Within this description, the system retains memory of the initial state throughout its evolution that is a property of non-ergodic behavior \cite{8}. Real quantum systems are open, which leads to fluctuations in the system and causes processes such as decoherence and information loss \cite{32,33}. These processes, in turn, play a key role in determining the ergodicity properties of quantum systems \cite{34,35}, which makes their study important not only from a fundamental but also from a practical point of view. With the development of quantum information theory, the study of questions about the reversibility of the time evolution of quantum systems has acquired great importance for the development of new technologies in the fields of quantum computing and quantum informatics \cite{zurek1984,barnum2002}. In particular, the study of these issues is important for solving problems related to the storage of information in quantum systems \cite{8}.

The stability of a quantum system is determined by the change in the time evolution of the wave function when perturbations are added. As a quantitative characteristic of the stability of a system, the Loschmidt echo is used, which is defined as $L(t) = \vert \langle \Psi'(t) \vert \Psi (t) \rangle \vert^{2}$ \cite{30,28,29,cucchietti2003,quan2006,jafari2017,yan2020}, where $\vert \Psi (t) \rangle$ and $\vert \Psi' (t) \rangle$ are wave functions of non-perturbed and perturbed systems, respectively. The Loschmidt echo is equal to the probability that at time $t$ the perturbed and non-perturbed systems are in the same state. If the Loschmidt echo remains close to $1$ for a long time, then the system retains memory of the initial state under the influence of external perturbations, and its behavior is non-ergodic \cite{30,28}. In the ergodic regime, the Loschmidt echo decays exponentially with time \cite{30,28}. This behavior means that the state of the perturbed system belongs to a subspace orthogonal to the state of the non-perturbed system. Thus, the transition from an ergodic behavior to a non-ergodic one is accompanied by the formation of stable states that can lead to a qualitative change in the properties of the system. 

An example of systems that can retain memory of the initial state and demonstrate non-ergodic behavior are discrete time crystals \cite{8,5,6,7}. The discrete time crystals are Hamiltonian systems subject to a periodic external force. The external periodic force generates a discrete time translation symmetry $H(t)=H(t+T)$ with respect to the time period of external driving, $T$ \cite{7,8}. If there exists a quantity in the system that changes with a multiple period of external driving $T'=mT, m=2,3,...$, then spontaneous breaking of time translation symmetry takes place \cite{7,8}. If, in addition to the existence of such a quantity, the system's behavior is stable to external perturbations, then it is called a discrete time crystal. Discrete time crystals are non-ergodic systems \cite{8,11,28,29,30} that, among other things, retain memory of their initial states and are time-reversible. There are a number of systems where it is possible to observe the time crystalline order, for example, driven atom-cavity systems \cite{9}, systems with trapped ions \cite{10}, NV-center systems \cite{11}, open optomechanical systems \cite{12} and NMR systems \cite{13,14}. The time crystalline order can also be observed in open systems with noisy environments, for example, in dissipative \cite{9,36,37,27,38,39,40} and boundary time crystals \cite{41,42,43,44,45}. 

The non-ergodic behavior in discrete time crystals is associated with the action of a periodic external force that causes periodic changes in the system \cite{8}. It has recently been shown that the spontaneous breaking of time translation symmetry can be observed in closed systems without periodic external driving \cite{15}. Such systems are described by the Schrödinger equation with a time-independent Hamiltonian. In such systems, there is no external force responsible for non-ergodic behavior in discrete time crystals. Therefore, it is a fundamental question whether such systems are discrete time crystals, which demonstrate non-ergodic behavior.

In this paper, we consider an autonomous quantum system consisting of two coupled single-mode cavities and the ring resonator of a finite length. Using the concept of the Loschmidt echo, we study the stability of the system dynamics to perturbation of the Hamiltonian parameters. We demonstrate that in such a system can occur a transition from the ergodic behavior to the non-ergodic one. This transition can be observed, for example, when the coupling strength between the single-mode cavities changes. In the region of the ergodic behavior, the Loschmidt echo decays with time for all initial conditions. This indicates the instability of the system's states to perturbations and the irreversibility in time of the system's dynamics. In the region of the non-ergodic behavior, there exists a special state that is perturbation-resistant and "infinitely living." The special state has a time crystalline order, i.e., it oscillates with a period twice as large as the timescale of the problem, determined by the time of one round trip of the resonator. We demonstrate that for the state with the time crystalline order, the Loschmidt echo oscillates over time, which is an indication of the non-ergodic behavior and memory preservation of the initial state even under small perturbations. Thus, we demonstrate the formation of a new stable phase state with broken time translation symmetry. We can conclude that the system under consideration is a new class of discrete time crystals - an autonomous discrete time crystal.

We demonstrate that at the point of transition from the autonomous discrete time crystal to the normal state, the time-averaged variance of the number of photons reaches a maximum. This behavior indicates an increase in the magnitude of fluctuations in the number of photons. This is a characteristic behavior of fluctuations for systems experiencing a transition with spontaneous symmetry breaking. Therefore, the variance of the number of photons can serve as a signature of the transition.

\section*{Model}
We consider a system consisting of two coupled single-mode cavities, one of which interacts with a resonator of finite length, $l$. We consider that the frequencies of both single-mode cavities are equal to $\omega_0$. The resonator is described as a set of modes, whose frequencies have the form $\omega_j = (\omega_0 + j \delta\omega)\left({1 + \varepsilon}\right)$, where $\delta\omega = 2 \pi c/l$ is a step between the modes' frequencies and $\varepsilon$ describes a perturbation of the mode's frequencies, which, for example, can appear due to thermal expansion of the resonator. Hereinafter, we will consider the time evolution of non-perturbed ($\varepsilon = 0$) and perturbed ($\varepsilon \ne 0$) systems. 

We use the following Hamiltonian for the system \cite{23}:

\begin{equation}
\begin{array}{l}
\hat H = {\omega _0}{\hat a_1^\dag} {{\hat a_1}} + {\omega _0}{\hat a_2^\dag} {{\hat a_2}} + \Omega (\hat a_1 {{\hat a_2^{\dag} }} + \hat a_2^{\dag} {{\hat a_1}}) + \\
\sum\limits_{j = -N/2}^{N/2} {{\omega_{j} \hat b_j^\dag {{\hat b}_j}} + \sum\limits_{j = -N/2}^{N/2} {{g_j}(\hat a_2 {{\hat b}_j^{\dag}} + \hat a_2^{\dag} {\hat b}_j})}
\end{array}
\label{eq:1}
\end{equation}
Here $\hat a_{1,2}$ and ${\hat a_{1,2}^{\dag}}$ are annihilation and creation operators of the first and second single-mode cavities, respectively. $\hat b_j$, $\hat b_j^{\dag}$ are annihilation and creation operators of the resonator's. The operators obey the bosonic commutation relations $[\hat a_{1,2}, {\hat a_{1,2}^{\dag}}] = 1$, $[\hat b_i, \hat b_j^{\dag}] = \delta_{ij}$. $\Omega$ is the coupling strength between the single-mode cavities. We consider the resonator's modes to interact with the second single-mode cavity with the same coupling strength, i.e., $g_j = g$ for all $j$. $N$ is a number of the ring resonator modes taken into account.

The dynamics of the system are determined by the time-dependent Schr\"{o}dinger equation for a wave function $\vert\Psi (t)\rangle$. The system is assumed to have only one excitation quantum that can be in one of the single-mode cavities or in one of the resonator modes. We look for the wave function in the following form \cite{23}:

\begin{equation}
\begin{array}{l}
\vert \Psi(t)\rangle = C_{a_1}(t)\vert 1,0,0\rangle + C_{a_2}(t)\vert 0,1,0\rangle + \\
\sum\limits_{j = -N/2}^{N/2} {C_{j}(t)\vert 0,0,1_{j}\rangle} 
\end{array}
\label{eq:2}
\end{equation}
where $\vert 1,0,0 \rangle$, $\vert 0,1,0 \rangle$, $\vert 0,0,1_{j}\rangle$ are the states, in which the excitation quantum is either in the first single-mode cavity, or in the second single-mode cavity, or in one of the resonator's modes, respectively. $C_{a_1}(t)$, $C_{a_2}(t)$, $C_{j}(t)$ are the amplitudes of probability of finding the excitation quantum in the corresponding states. In what follows, we denote by $\vert \Psi(t)\rangle$ the wave function of the unperturbed system ($\varepsilon = 0$) and by $\vert \Psi'(t)\rangle$ the wave function of the perturbed system ($\varepsilon \ne 0$).

\section*{Special state in the system under consideration}
\begin{figure}[htbp]
\centering\includegraphics[width=\linewidth]{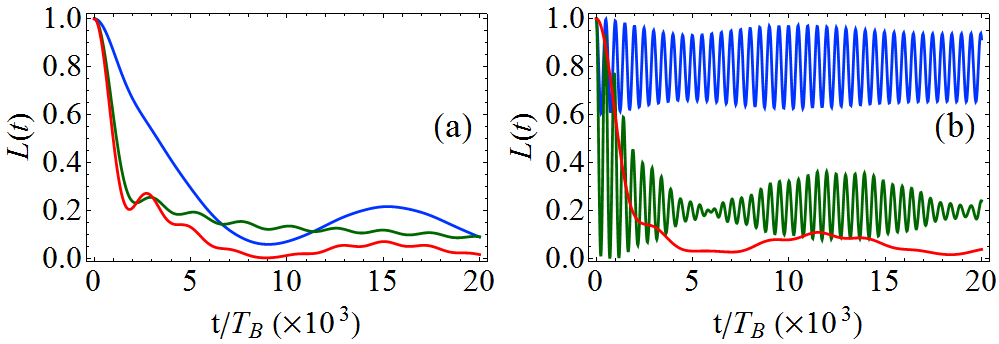}
\caption{Time dependence of the Loschmidt echo when $\Omega = 10 \Omega_{TC}$ (a) and $\Omega = 0.5 \Omega_{TC}$ (b). The initial state is ${C_{{a_{\,1}}}} = 1$, ${C_{{a_{\,2}}}} = 0$ (the blue lines); ${C_{{a_{\,1}}}} = 1/\sqrt 2$, ${C_{{a_{\,2}}}} = 1/\sqrt 2$ (the green lines); ${C_{{a_{\,1}}}} = 0$, ${C_{{a_{\,2}}}} = 1$ (the red lines), ${C_j} = 0$. Here $T_B$ is the time of one bypass of the resonator, $N=50$, $\delta\omega = 4\cdot 10^{-3} \omega_0$, $g=6\cdot 10^{-3}\omega_0$, $\Omega_{TC}=g/\sqrt{2}$, $\varepsilon = 10^{-5}$.}
\label{fig1}
\end{figure}

\begin{figure}[htbp]
\centering\includegraphics[width=0.8\linewidth]{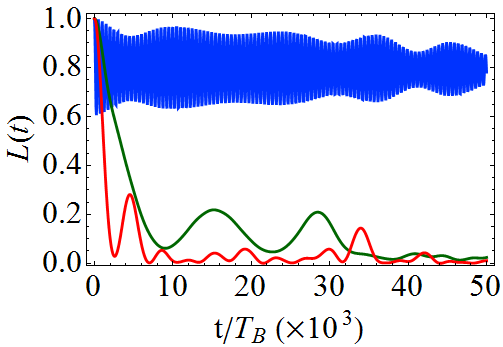}
\caption{Time dependence of the Loschmidt echo when $\Omega = 0.5 \Omega_{TC}$ (the blue line), $\Omega = 5 \Omega_{TC}$ (the green line), $\Omega = 10 \Omega_{TC}$ (the red line). The initial state is ${C_{{a_{\,1}}}} = 1$, ${C_{{a_{\,2}}}} = 0$, ${C_j} = 0$. Here $N=50$, $\delta\omega = 4\cdot 10^{-3} \omega_0$, $g=6\cdot 10^{-3}\omega_0$, $\Omega_{TC}=g/\sqrt{2}$, $\varepsilon = 10^{-5}$.}
\label{fig2}
\end{figure}

We investigate the stability of the system dynamics to perturbations. We find the time dependence of the wave function for the non-perturbed ($\varepsilon =0$) and perturbed system ($\varepsilon \ne 0$). In order to characterize the stability of the system to perturbations, we use the Loschmidt echo \cite{29,cucchietti2003,quan2006,jafari2017,yan2020} ($L(t) = \vert \langle \Psi'(t) \vert \Psi (t) \rangle \vert^{2}$). Our calculations show that in the system under consideration, the temporal dynamics of the Loschmidt echo qualitatively depend on the coupling constant between the single-mode resonators, $\Omega $, and the type of initial state. When $\Omega  >  > \Omega_{TC}=g/\sqrt{2}$, at large times for any initial states, the average value of the Loschmidt echo takes values much smaller than $1$ [Figure~\ref{fig1}a]. When $\Omega  \sim \Omega_{TC}$, there is a special initial state, which is defined as ${C_{{a_{\,1}}}}\left( 0 \right) = 1$, ${C_{{a_{\,2}}}}\left( 0 \right) = 0$, ${C_j}\left( 0 \right) = 0$. If the initial state coincides with the special state, the Loschmidt echo oscillates without decaying [Figure~\ref{fig1}b]. If the initial state is orthogonal to the special state, then at large times the Loschmidt echo takes values much smaller than $1$ [Figure~\ref{fig1}b]. The average value of the Loschmidt echo in this case is equal to $N^{-1}$ \cite{29,30}, where $N$ is the number of degrees of freedom in the system. If the initial state is not orthogonal to the special state, then at large times the Loschmidt echo oscillates near a value proportional to the contribution of the special state to the initial state [Figure~\ref{fig1}b]. Thus, for the special initial state, the time dynamics of the Loschmidt echo exhibit a transition with increasing $\Omega $, changing from periodic oscillations near a value close to $1$ to quasi-random fluctuations near $0$ [Figure~\ref{fig2}]. 

To better study the properties of the special state, we calculate the value $p\left( t \right) = {\left| {\left\langle {{\Psi \left( 0 \right)}}
 \mathrel{\left | {\vphantom {{\Psi \left( 0 \right)} {\Psi \left( t \right)}}}
 \right. \kern-\nulldelimiterspace}
 {{\Psi \left( t \right)}} \right\rangle } \right|^2}$, which shows the probability of the system returning to the initial state at time $t$.

\begin{figure}[htbp]
\centering\includegraphics[width=0.8\linewidth]{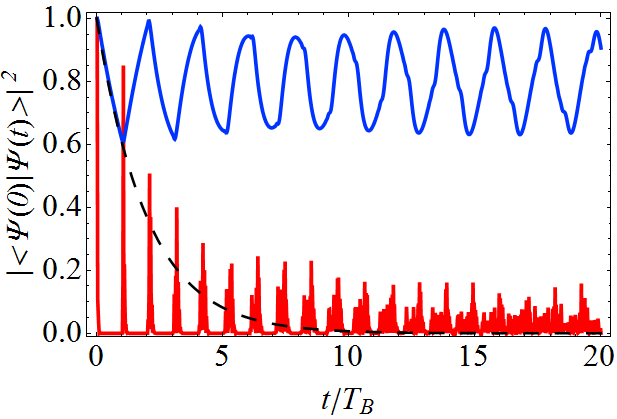}
\caption{Time dependence of $p\left( t \right) = {\left| {\left\langle {{\Psi \left( 0 \right)}}
 \mathrel{\left | {\vphantom {{\Psi \left( 0 \right)} {\Psi \left( t \right)}}}
 \right. \kern-\nulldelimiterspace}
 {{\Psi \left( t \right)}} \right\rangle } \right|^2}$ when $\Omega = 0.5 \Omega_{TC}$ (the blue line), $\Omega = 10 \Omega_{TC}$ (the red line). The initial state is ${C_{{a_{\,1}}}} = 1$, ${C_{{a_{\,2}}}} = 0$, ${C_j} = 0$. $N=50$, $\delta\omega = 4\cdot 10^{-3} \omega_0$, $g=6\cdot 10^{-3}\omega_0$, $\Omega_{TC}=g/\sqrt{2}$. The black dashed line shows the dependence $\exp \left( { - \gamma t} \right)$, where $\gamma  = \frac{{{2\Omega ^2}}}{{\pi {g^2}}}\delta \omega$ and $\Omega = 0.5 \Omega_{TC}$.}
\label{fig3}
\end{figure}

In the system under consideration, there is a timescale that is determined by the ring resonator bypass time ${T_B} = l/c = 2\pi /\delta \omega $. Therefore, one can expect that the probability of the system returning to the initial state will demonstrate maxima with a period approximately equal to ${T_B}$. For states orthogonal to the special state, such behavior is observed for all values of $\Omega $. For the special state, similar behavior is observed at $\Omega  >  > \Omega_{TC}$ [Figure~\ref{fig3}]. At $\Omega  < \Omega_{TC}$, for the special initial state, the system demonstrates the regime with a doubled period. That is, $p\left( t \right)$ demonstrates maxima with a period equal to $2{T_B}$ [Figure~\ref{fig3}]. In this regime, the wave function of the system remains close to the initial state at all times [Figure~\ref{fig3}].

For the special state $p\left( t \right) = {\left| {{C_{{a_1}}}\left( t \right)} \right|^2}$. Therefore, it is clear that oscillations of $p\left( t \right)$ near $1$ correspond to the fact that the excitation quantum remains in the first single-mode cavity. At coupling strengths $\Omega < \Omega_{TC}$ the state cannot completely decay after the time $T_B$ and then the reverse flow restores the state at the time $2 T_B$ [Figure~\ref{fig3}]. Thus, there is an "infinitely living" state in the system, and, moreover, this state demonstrates broken discrete time symmetry. The existence of such a state is due to the periodic change in the direction of the flow between the first single-mode cavity and the ring resonator. Indeed, the rate of excitation transfer from the first single-mode cavity to the rest of the system can be estimated as $\gamma  = \frac{{{2\Omega ^2}}}{{\pi {g^2}}}\delta \omega$ (see Supplementary Materials). This expression correctly determines the rate of excitation transfer from the first single-mode cavity to the rest of the system at small times [Figure~\ref{fig3}]. However, at subsequent times, the interaction of the system with the single-mode cavity leads to the reverse transition, which keeps the system in a state close to the initial one. As a result, the lifetime of the system near the special initial state turns out to be much longer than ${\gamma ^{ - 1}}$ (our calculations show that the system remains near the special initial state for any simulation time). The change in flow direction occurs after a time interval $T_B$ and the full period of the system is $2 T_B$. Thus, at $\Omega  < \Omega_{TC}$, there is an "infinitely living" state in the system, which is characterized by broken time symmetry. Our calculations of the Loschmidt echo show that the state with time crystal order is stable to perturbations [Figures~\ref{fig1},\ref{fig2}].

\section*{Autonomous discrete time crystal}
The dynamics of the Loschmidt echo make it possible to determine not only the stability of the system to perturbations, but also allow one to determine whether the system is ergodic and time-reversible. The Schrödinger equation is time-reversible. Therefore, a quantum system must retain memory of the initial state (knowing the current state, we can always determine the initial state). However, in a real system, there are always some perturbations (for example, caused by thermal fluctuations), which change the wave function. If the state of the system is not stable to perturbations, then at large times it becomes impossible to restore the initial state of the system. In other words, the evolution of the system is irreversible in time. In this case, the Loschmidt echo exponentially decays in time, which is a criterion for the ergodicity of the system \cite{28}. If the state of the system is stable to perturbations, then even at large times it is possible to restore the initial state of the system. In this case, the evolution of the system is reversible in time, and the system is non-ergodic. The undamped evolution of the Loschmidt echo is an indication of the non-ergodicity of the system \cite{28}.

Our calculations show that in the system under consideration there is a special state that is "infinitely living" [Figure~\ref{fig3}] and stable to perturbations [Figures~\ref{fig1},\ref{fig2}]. This state has a time crystalline order and exists when $\Omega  < \Omega_{TC}$. As $\Omega$ increases, all states become unstable to perturbations. Thus, we can conclude that the changing of the coupling strength $\Omega$ leads to a transition from a non-ergodic system to an ergodic one. With such a transition, the dynamics of the system become irreversible in time for all initial states. The obtained results allow us to conclude that when $\Omega  < \Omega_{TC}$ a new phase state occurs in the system. The system in this regime is a new type of discrete time crystal, namely, an autonomous discrete time crystal.

\section*{Maximum variance of the number of photons as a criterion for the transition to the autonomous discrete-time crystal regime}

\begin{figure}[htbp]
\centering\includegraphics[width=0.8\linewidth]{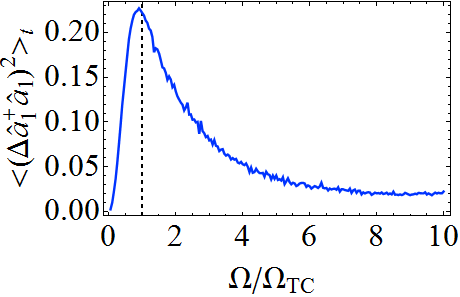}
\caption{Dependence of the time-averaged variance of the number of photons in the first single-mode cavity for the special state on $\Omega$.}
\label{fig5}
\end{figure}

The transition point from the autonomous discrete time crystal to the normal state can be detected using the variance of the number of photons in the first single-mode cavity, $\Delta {\left( {\hat a_1^\dag {{\hat a}_1}} \right)^2}$:

\begin{equation}
\Delta {\left( {\hat a_1^\dag {{\hat a}_1}} \right)^2} = {{{\left| {{C_{{a_1}}}\left( t \right)} \right|}^2} - {{\left| {{C_{{a_1}}}\left( t \right)} \right|}^4}}
\label{eq:3}
\end{equation}
The number of photons in the first single-mode cavity depends on time [Figure~\ref{fig3}]. For the special state, such a quantity changes in time with a characteristic time scale $2 T_B$ in the autonomous time crystal mode and $T_B$ in the normal mode. Therefore, the variance of the number of photons also depends on time. As a quantity characterizing the transition in the system, we use the variance averaged over a large time interval:

\begin{equation}
{\left\langle \Delta {\left( {\hat a_1^\dag {{\hat a}_1}} \right)^2}  \right\rangle _t} = \int\limits_{{T_0}}^{{T_0} + m{T_B}} {\Delta {\left( {\hat a_1^\dag {{\hat a}_1}} \right)^2} dt}
\label{eq:4}
\end{equation}
Here ${T_0} >  > {T_B}$, $m$ is an integer and $m> >1$.
Numerical calculation shows that for a special state, the time-averaged variance of the number of photons in the first single-mode cavity has a maximum near the transition point from the regime of discrete time crystal to the normal state [Figure~\ref{fig5}]. At the same time, for a state orthogonal to the special state, the variance of the number of photons has no maximum.

The variance behavior indicates an increase in the magnitude of fluctuations in the number of photons near the point of transition from the autonomous discrete time crystal to the normal state. The similar behavior of fluctuations is characteristic of transitions caused by spontaneous symmetry breaking \cite{34}. Thus, the maximum of time-averaged variance of the number of photons can serve as a signature of the transition to the autonomous discrete time crystal regime.

\section*{Transition from non-ergodic system to ergodic one caused by changing the size of the resonator}
The transition from the non-ergodic system to the ergodic one can occur not only from the change of the coupling strength between single-mode resonators but, for example, from the change of the resonator's length. The length of the resonator determines the step between the modes’ frequencies ($\delta \omega  = 2\pi /l$) and the interaction constant between the second single-mode resonator and the resonator modes ($g \sim {l^{ - 1/2}}$). This dependence on the length of the resonator is due to the fact that $g$ is proportional to the amplitude of the electric field in the resonator mode per quantum \cite{LL4}, which, according to the second quantization procedure, is proportional to ${l^{ - 1/2}}$ \cite{LL4}.

\begin{figure}[htbp]
\centering\includegraphics[width=0.8\linewidth]{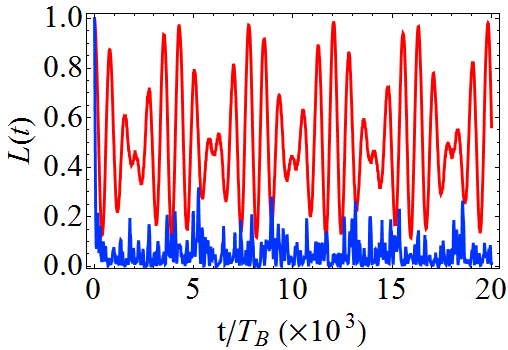}
\caption{Time dependence of the Loschmidt echo when  $L = 250{\lambda _0}$ (the blue line) and when $L = 10{\lambda _0}$ (the red line), where ${\lambda _0} = \frac{{2\pi c}}{{{\omega _0}}}$. In the first case 
$\delta\omega = 4\cdot 10^{-3} \omega_0$ and $g=6\cdot 10^{-3}\omega_0$; in the second case $\delta\omega = 10^{-1} \omega_0$ and $g=3\cdot 10^{-2}\omega_0$. The initial state is ${C_{{a_{\,1}}}} = 1$, ${C_{{a_{\,2}}}} = 0$, ${C_j} = 0$. $\Omega_{TC}=g/\sqrt{2}$, $\varepsilon = 10^{-3}$.}
\label{fig4}
\end{figure}

The special state, which is 'infinitely living' and stable to perturbations, exists when $\Omega  < \Omega_{TC}$. Therefore, changing the size of the resonator can lead to a transition from a system in which the special state exists to a system in which such states are absent. The evolution of the Loschmidt echo shows that such a transition is indeed observed in the system [Figure~\ref{fig4}]. As the resonator size decreases, the exponential decay of the Loschmidt echo is replaced by periodic oscillations [Figure~\ref{fig4}]. Thus, the system undergoes the transition from ergodic to non-ergodic behavior, and the system changes from time-irreversible to time-reversible. Since the rate of excitation transfer from the first single-mode resonator to the rest of the system ($\gamma  = \frac{{{2\Omega ^2}}}{{\pi {g^2}}}\delta \omega$) does not depend on the size of the resonator ($\delta \omega  \sim {l^{ - 1}}$ and $g \sim {l^{ - 1/2}}$), we conclude that such a transition is not associated with a change in the magnitude of the interaction between the first single-mode resonator and the rest of the system.

To summarize, we can conclude that the transition from ergodic to non-ergodic behavior, which occurs with a change in the resonator size, is associated with a change in the number of degrees of freedom in the system. The system under consideration, consisting of three coupled resonators, is quite simple, not only from a theoretical point of view, but also convenient for implementation in practice. Such a simple system provides a convenient basis for studying transitions from time-reversible to time-irreversible behavior that occur with a change in the number of degrees of freedom.

\section*{Conclusion}
We consider the system of two coupled single-mode cavities, one of which interacts with the ring resonator. Observing the time evolution of the Loschmidt echo, we demonstrate that the system can experience the transition from the ergodic behavior to the non-ergodic one. This transition can be observed by changing the system's parameters, for example, coupling strength between the single-mode cavities. In the ergodic region, the Loschmidt echo decays with time for all initial conditions. This behavior indicates the instability of the system's states to perturbations and the irreversibility in time of the system's dynamics. In the non-ergodic region, there is a special state that is perturbation-resistant and "infinitely living." For such a state, the dynamics of the system are time-reversible. The transition to the non-ergodic behavior indicates the formation of a new phase state in the system. This gives grounds to consider the system under study as a new type of discrete time crystal - the autonomous discrete time crystal.

We demonstrate that at the point of transition from the autonomous discrete time crystal to the normal state, the time-averaged variance of the number of photons reaches a maximum. That makes it possible to use the variance of the number of photons as a signature of its transition.

We show that in the system under consideration, the transition from ergodic to non-ergodic behavior can be observed by changing the resonator length. We argue that such a transition is associated with a change in the number of degrees of freedom in the system. Thus, the system under consideration provides a simple framework for studying transitions from time-reversible to time-irreversible behavior with a change in the number of degrees of freedom.

\section*{Acknowledgment}
The study was financially supported by a Grant from Russian Science Foundation (Project No. 20-72-10057). E.S.A thanks foundation for the advancement of theoretical physics and mathematics 'Basis'.

\nocite{*}

\bibliography{apssamp}

\providecommand{\noopsort}[1]{}\providecommand{\singleletter}[1]{#1}%
\begin{thebibliography}{42}%
\makeatletter
\providecommand \@ifxundefined [1]{%
 \@ifx{#1\undefined}
}%
\providecommand \@ifnum [1]{%
 \ifnum #1\expandafter \@firstoftwo
 \else \expandafter \@secondoftwo
 \fi
}%
\providecommand \@ifx [1]{%
 \ifx #1\expandafter \@firstoftwo
 \else \expandafter \@secondoftwo
 \fi
}%
\providecommand \natexlab [1]{#1}%
\providecommand \enquote  [1]{``#1''}%
\providecommand \bibnamefont  [1]{#1}%
\providecommand \bibfnamefont [1]{#1}%
\providecommand \citenamefont [1]{#1}%
\providecommand \href@noop [0]{\@secondoftwo}%
\providecommand \href [0]{\begingroup \@sanitize@url \@href}%
\providecommand \@href[1]{\@@startlink{#1}\@@href}%
\providecommand \@@href[1]{\endgroup#1\@@endlink}%
\providecommand \@sanitize@url [0]{\catcode `\\12\catcode `\$12\catcode `\&12\catcode `\#12\catcode `\^12\catcode `\_12\catcode `\%12\relax}%
\providecommand \@@startlink[1]{}%
\providecommand \@@endlink[0]{}%
\providecommand \url  [0]{\begingroup\@sanitize@url \@url }%
\providecommand \@url [1]{\endgroup\@href {#1}{\urlprefix }}%
\providecommand \urlprefix  [0]{URL }%
\providecommand \Eprint [0]{\href }%
\providecommand \doibase [0]{https://doi.org/}%
\providecommand \selectlanguage [0]{\@gobble}%
\providecommand \bibinfo  [0]{\@secondoftwo}%
\providecommand \bibfield  [0]{\@secondoftwo}%
\providecommand \translation [1]{[#1]}%
\providecommand \BibitemOpen [0]{}%
\providecommand \bibitemStop [0]{}%
\providecommand \bibitemNoStop [0]{.\EOS\space}%
\providecommand \EOS [0]{\spacefactor3000\relax}%
\providecommand \BibitemShut  [1]{\csname bibitem#1\endcsname}%
\let\auto@bib@innerbib\@empty
\bibitem [{\citenamefont {Landau}\ and\ \citenamefont {Lifshitz}(2013{\natexlab{a}})}]{31}%
  \BibitemOpen
  \bibfield  {author} {\bibinfo {author} {\bibfnamefont {L.~D.}\ \bibnamefont {Landau}}\ and\ \bibinfo {author} {\bibfnamefont {E.~M.}\ \bibnamefont {Lifshitz}},\ }\href@noop {} {\emph {\bibinfo {title} {Quantum mechanics: non-relativistic theory}}},\ Vol.~\bibinfo {volume} {3}\ (\bibinfo  {publisher} {Elsevier},\ \bibinfo {year} {2013})\BibitemShut {NoStop}%
\bibitem [{\citenamefont {Zaletel}\ \emph {et~al.}(2023)\citenamefont {Zaletel}, \citenamefont {Lukin}, \citenamefont {Monroe}, \citenamefont {Nayak}, \citenamefont {Wilczek},\ and\ \citenamefont {Yao}}]{8}%
  \BibitemOpen
  \bibfield  {author} {\bibinfo {author} {\bibfnamefont {M.~P.}\ \bibnamefont {Zaletel}}, \bibinfo {author} {\bibfnamefont {M.}~\bibnamefont {Lukin}}, \bibinfo {author} {\bibfnamefont {C.}~\bibnamefont {Monroe}}, \bibinfo {author} {\bibfnamefont {C.}~\bibnamefont {Nayak}}, \bibinfo {author} {\bibfnamefont {F.}~\bibnamefont {Wilczek}},\ and\ \bibinfo {author} {\bibfnamefont {N.~Y.}\ \bibnamefont {Yao}},\ }\bibfield  {title} {\bibinfo {title} {Colloquium: Quantum and classical discrete time crystals},\ }\href@noop {} {\bibfield  {journal} {\bibinfo  {journal} {Rev. Mod. Phys.}\ }\textbf {\bibinfo {volume} {95}},\ \bibinfo {pages} {031001} (\bibinfo {year} {2023})}\BibitemShut {NoStop}%
\bibitem [{\citenamefont {Zeh}(1970)}]{32}%
  \BibitemOpen
  \bibfield  {author} {\bibinfo {author} {\bibfnamefont {H.~D.}\ \bibnamefont {Zeh}},\ }\bibfield  {title} {\bibinfo {title} {On the interpretation of measurement in quantum theory},\ }\href@noop {} {\bibfield  {journal} {\bibinfo  {journal} {Found. Phys.}\ }\textbf {\bibinfo {volume} {1}},\ \bibinfo {pages} {69} (\bibinfo {year} {1970})}\BibitemShut {NoStop}%
\bibitem [{\citenamefont {Schlosshauer}(2019)}]{33}%
  \BibitemOpen
  \bibfield  {author} {\bibinfo {author} {\bibfnamefont {M.}~\bibnamefont {Schlosshauer}},\ }\bibfield  {title} {\bibinfo {title} {Quantum decoherence},\ }\href@noop {} {\bibfield  {journal} {\bibinfo  {journal} {Phys. Rep.}\ }\textbf {\bibinfo {volume} {831}},\ \bibinfo {pages} {1} (\bibinfo {year} {2019})}\BibitemShut {NoStop}%
\bibitem [{\citenamefont {Landau}\ and\ \citenamefont {Lifshitz}(2013{\natexlab{b}})}]{34}%
  \BibitemOpen
  \bibfield  {author} {\bibinfo {author} {\bibfnamefont {L.~D.}\ \bibnamefont {Landau}}\ and\ \bibinfo {author} {\bibfnamefont {E.~M.}\ \bibnamefont {Lifshitz}},\ }\href@noop {} {\emph {\bibinfo {title} {Statistical Physics: Volume 5}}},\ Vol.~\bibinfo {volume} {5}\ (\bibinfo  {publisher} {Elsevier},\ \bibinfo {year} {2013})\BibitemShut {NoStop}%
\bibitem [{\citenamefont {Vinjanampathy}\ and\ \citenamefont {Anders}(2016)}]{35}%
  \BibitemOpen
  \bibfield  {author} {\bibinfo {author} {\bibfnamefont {S.}~\bibnamefont {Vinjanampathy}}\ and\ \bibinfo {author} {\bibfnamefont {J.}~\bibnamefont {Anders}},\ }\bibfield  {title} {\bibinfo {title} {Quantum thermodynamics},\ }\href@noop {} {\bibfield  {journal} {\bibinfo  {journal} {Contemp. Phys.}\ }\textbf {\bibinfo {volume} {57}},\ \bibinfo {pages} {545} (\bibinfo {year} {2016})}\BibitemShut {NoStop}%
\bibitem [{\citenamefont {Zurek}(1984)}]{zurek1984}%
  \BibitemOpen
  \bibfield  {author} {\bibinfo {author} {\bibfnamefont {W.~H.}\ \bibnamefont {Zurek}},\ }\bibfield  {title} {\bibinfo {title} {Reversibility and stability of information processing systems},\ }\href@noop {} {\bibfield  {journal} {\bibinfo  {journal} {Phys. Rev. Lett.}\ }\textbf {\bibinfo {volume} {53}},\ \bibinfo {pages} {391} (\bibinfo {year} {1984})}\BibitemShut {NoStop}%
\bibitem [{\citenamefont {Barnum}\ and\ \citenamefont {Knill}(2002)}]{barnum2002}%
  \BibitemOpen
  \bibfield  {author} {\bibinfo {author} {\bibfnamefont {H.}~\bibnamefont {Barnum}}\ and\ \bibinfo {author} {\bibfnamefont {E.}~\bibnamefont {Knill}},\ }\bibfield  {title} {\bibinfo {title} {Reversing quantum dynamics with near-optimal quantum and classical fidelity},\ }\href@noop {} {\bibfield  {journal} {\bibinfo  {journal} {J. Math. Phys.}\ }\textbf {\bibinfo {volume} {43}},\ \bibinfo {pages} {2097} (\bibinfo {year} {2002})}\BibitemShut {NoStop}%
\bibitem [{\citenamefont {Goussev}\ \emph {et~al.}(2012)\citenamefont {Goussev}, \citenamefont {Jalabert}, \citenamefont {Pastawski},\ and\ \citenamefont {Wisniacki}}]{30}%
  \BibitemOpen
  \bibfield  {author} {\bibinfo {author} {\bibfnamefont {A.}~\bibnamefont {Goussev}}, \bibinfo {author} {\bibfnamefont {R.~A.}\ \bibnamefont {Jalabert}}, \bibinfo {author} {\bibfnamefont {H.~M.}\ \bibnamefont {Pastawski}},\ and\ \bibinfo {author} {\bibfnamefont {D.}~\bibnamefont {Wisniacki}},\ }\bibfield  {title} {\bibinfo {title} {Loschmidt echo},\ }\href@noop {} {\bibfield  {journal} {\bibinfo  {journal} {arXiv preprint arXiv:1206.6348}\ } (\bibinfo {year} {2012})}\BibitemShut {NoStop}%
\bibitem [{\citenamefont {Serbyn}\ and\ \citenamefont {Abanin}(2017)}]{28}%
  \BibitemOpen
  \bibfield  {author} {\bibinfo {author} {\bibfnamefont {M.}~\bibnamefont {Serbyn}}\ and\ \bibinfo {author} {\bibfnamefont {D.~A.}\ \bibnamefont {Abanin}},\ }\bibfield  {title} {\bibinfo {title} {Loschmidt echo in many-body localized phases},\ }\href@noop {} {\bibfield  {journal} {\bibinfo  {journal} {Phys. Rev. B}\ }\textbf {\bibinfo {volume} {96}},\ \bibinfo {pages} {014202} (\bibinfo {year} {2017})}\BibitemShut {NoStop}%
\bibitem [{\citenamefont {Gorin}\ \emph {et~al.}(2006)\citenamefont {Gorin}, \citenamefont {Prosen}, \citenamefont {Seligman},\ and\ \citenamefont {{\v{Z}}nidari{\v{c}}}}]{29}%
  \BibitemOpen
  \bibfield  {author} {\bibinfo {author} {\bibfnamefont {T.}~\bibnamefont {Gorin}}, \bibinfo {author} {\bibfnamefont {T.}~\bibnamefont {Prosen}}, \bibinfo {author} {\bibfnamefont {T.~H.}\ \bibnamefont {Seligman}},\ and\ \bibinfo {author} {\bibfnamefont {M.}~\bibnamefont {{\v{Z}}nidari{\v{c}}}},\ }\bibfield  {title} {\bibinfo {title} {Dynamics of loschmidt echoes and fidelity decay},\ }\href@noop {} {\bibfield  {journal} {\bibinfo  {journal} {Phys. Rep.}\ }\textbf {\bibinfo {volume} {435}},\ \bibinfo {pages} {33} (\bibinfo {year} {2006})}\BibitemShut {NoStop}%
\bibitem [{\citenamefont {Cucchietti}\ \emph {et~al.}(2003)\citenamefont {Cucchietti}, \citenamefont {Dalvit}, \citenamefont {Paz},\ and\ \citenamefont {Zurek}}]{cucchietti2003}%
  \BibitemOpen
  \bibfield  {author} {\bibinfo {author} {\bibfnamefont {F.~M.}\ \bibnamefont {Cucchietti}}, \bibinfo {author} {\bibfnamefont {D.~A.}\ \bibnamefont {Dalvit}}, \bibinfo {author} {\bibfnamefont {J.~P.}\ \bibnamefont {Paz}},\ and\ \bibinfo {author} {\bibfnamefont {W.~H.}\ \bibnamefont {Zurek}},\ }\bibfield  {title} {\bibinfo {title} {Decoherence and the loschmidt echo},\ }\href@noop {} {\bibfield  {journal} {\bibinfo  {journal} {Phys. Rev. Lett.}\ }\textbf {\bibinfo {volume} {91}},\ \bibinfo {pages} {210403} (\bibinfo {year} {2003})}\BibitemShut {NoStop}%
\bibitem [{\citenamefont {Quan}\ \emph {et~al.}(2006)\citenamefont {Quan}, \citenamefont {Song}, \citenamefont {Liu}, \citenamefont {Zanardi},\ and\ \citenamefont {Sun}}]{quan2006}%
  \BibitemOpen
  \bibfield  {author} {\bibinfo {author} {\bibfnamefont {H.}~\bibnamefont {Quan}}, \bibinfo {author} {\bibfnamefont {Z.}~\bibnamefont {Song}}, \bibinfo {author} {\bibfnamefont {X.~F.}\ \bibnamefont {Liu}}, \bibinfo {author} {\bibfnamefont {P.}~\bibnamefont {Zanardi}},\ and\ \bibinfo {author} {\bibfnamefont {C.-P.}\ \bibnamefont {Sun}},\ }\bibfield  {title} {\bibinfo {title} {Decay of loschmidt echo enhanced by quantum criticality},\ }\href@noop {} {\bibfield  {journal} {\bibinfo  {journal} {Phys. Rev. Lett.}\ }\textbf {\bibinfo {volume} {96}},\ \bibinfo {pages} {140604} (\bibinfo {year} {2006})}\BibitemShut {NoStop}%
\bibitem [{\citenamefont {Jafari}\ and\ \citenamefont {Johannesson}(2017)}]{jafari2017}%
  \BibitemOpen
  \bibfield  {author} {\bibinfo {author} {\bibfnamefont {R.}~\bibnamefont {Jafari}}\ and\ \bibinfo {author} {\bibfnamefont {H.}~\bibnamefont {Johannesson}},\ }\bibfield  {title} {\bibinfo {title} {Loschmidt echo revivals: Critical and noncritical},\ }\href@noop {} {\bibfield  {journal} {\bibinfo  {journal} {Phys. Rev. Let.}\ }\textbf {\bibinfo {volume} {118}},\ \bibinfo {pages} {015701} (\bibinfo {year} {2017})}\BibitemShut {NoStop}%
\bibitem [{\citenamefont {Yan}\ \emph {et~al.}(2020)\citenamefont {Yan}, \citenamefont {Cincio},\ and\ \citenamefont {Zurek}}]{yan2020}%
  \BibitemOpen
  \bibfield  {author} {\bibinfo {author} {\bibfnamefont {B.}~\bibnamefont {Yan}}, \bibinfo {author} {\bibfnamefont {L.}~\bibnamefont {Cincio}},\ and\ \bibinfo {author} {\bibfnamefont {W.~H.}\ \bibnamefont {Zurek}},\ }\bibfield  {title} {\bibinfo {title} {Information scrambling and loschmidt echo},\ }\href@noop {} {\bibfield  {journal} {\bibinfo  {journal} {Phys. Rev. Lett.}\ }\textbf {\bibinfo {volume} {124}},\ \bibinfo {pages} {160603} (\bibinfo {year} {2020})}\BibitemShut {NoStop}%
\bibitem [{\citenamefont {Yao}\ \emph {et~al.}(2017)\citenamefont {Yao}, \citenamefont {Potter}, \citenamefont {Potirniche},\ and\ \citenamefont {Vishwanath}}]{5}%
  \BibitemOpen
  \bibfield  {author} {\bibinfo {author} {\bibfnamefont {N.~Y.}\ \bibnamefont {Yao}}, \bibinfo {author} {\bibfnamefont {A.~C.}\ \bibnamefont {Potter}}, \bibinfo {author} {\bibfnamefont {I.-D.}\ \bibnamefont {Potirniche}},\ and\ \bibinfo {author} {\bibfnamefont {A.}~\bibnamefont {Vishwanath}},\ }\bibfield  {title} {\bibinfo {title} {Discrete time crystals: Rigidity, criticality, and realizations},\ }\href@noop {} {\bibfield  {journal} {\bibinfo  {journal} {Phys. Rev. Lett.}\ }\textbf {\bibinfo {volume} {118}},\ \bibinfo {pages} {030401} (\bibinfo {year} {2017})}\BibitemShut {NoStop}%
\bibitem [{\citenamefont {Gong}\ \emph {et~al.}(2018)\citenamefont {Gong}, \citenamefont {Hamazaki},\ and\ \citenamefont {Ueda}}]{6}%
  \BibitemOpen
  \bibfield  {author} {\bibinfo {author} {\bibfnamefont {Z.}~\bibnamefont {Gong}}, \bibinfo {author} {\bibfnamefont {R.}~\bibnamefont {Hamazaki}},\ and\ \bibinfo {author} {\bibfnamefont {M.}~\bibnamefont {Ueda}},\ }\bibfield  {title} {\bibinfo {title} {Discrete time-crystalline order in cavity and circuit qed systems},\ }\href@noop {} {\bibfield  {journal} {\bibinfo  {journal} {Phys. Rev. Lett.}\ }\textbf {\bibinfo {volume} {120}},\ \bibinfo {pages} {040404} (\bibinfo {year} {2018})}\BibitemShut {NoStop}%
\bibitem [{\citenamefont {Else}\ \emph {et~al.}(2020)\citenamefont {Else}, \citenamefont {Monroe}, \citenamefont {Nayak},\ and\ \citenamefont {Yao}}]{7}%
  \BibitemOpen
  \bibfield  {author} {\bibinfo {author} {\bibfnamefont {D.~V.}\ \bibnamefont {Else}}, \bibinfo {author} {\bibfnamefont {C.}~\bibnamefont {Monroe}}, \bibinfo {author} {\bibfnamefont {C.}~\bibnamefont {Nayak}},\ and\ \bibinfo {author} {\bibfnamefont {N.~Y.}\ \bibnamefont {Yao}},\ }\bibfield  {title} {\bibinfo {title} {Discrete time crystals},\ }\href@noop {} {\bibfield  {journal} {\bibinfo  {journal} {Annu. Rev. Condens. Matter Phys.}\ }\textbf {\bibinfo {volume} {11}},\ \bibinfo {pages} {467} (\bibinfo {year} {2020})}\BibitemShut {NoStop}%
\bibitem [{\citenamefont {Choi}\ \emph {et~al.}(2017)\citenamefont {Choi}, \citenamefont {Choi}, \citenamefont {Landig}, \citenamefont {Kucsko}, \citenamefont {Zhou}, \citenamefont {Isoya}, \citenamefont {Jelezko}, \citenamefont {Onoda}, \citenamefont {Sumiya}, \citenamefont {Khemani} \emph {et~al.}}]{11}%
  \BibitemOpen
  \bibfield  {author} {\bibinfo {author} {\bibfnamefont {S.}~\bibnamefont {Choi}}, \bibinfo {author} {\bibfnamefont {J.}~\bibnamefont {Choi}}, \bibinfo {author} {\bibfnamefont {R.}~\bibnamefont {Landig}}, \bibinfo {author} {\bibfnamefont {G.}~\bibnamefont {Kucsko}}, \bibinfo {author} {\bibfnamefont {H.}~\bibnamefont {Zhou}}, \bibinfo {author} {\bibfnamefont {J.}~\bibnamefont {Isoya}}, \bibinfo {author} {\bibfnamefont {F.}~\bibnamefont {Jelezko}}, \bibinfo {author} {\bibfnamefont {S.}~\bibnamefont {Onoda}}, \bibinfo {author} {\bibfnamefont {H.}~\bibnamefont {Sumiya}}, \bibinfo {author} {\bibfnamefont {V.}~\bibnamefont {Khemani}}, \emph {et~al.},\ }\bibfield  {title} {\bibinfo {title} {Observation of discrete time-crystalline order in a disordered dipolar many-body system},\ }\href@noop {} {\bibfield  {journal} {\bibinfo  {journal} {Nature}\ }\textbf {\bibinfo {volume} {543}},\ \bibinfo {pages} {221} (\bibinfo {year} {2017})}\BibitemShut {NoStop}%
\bibitem [{\citenamefont {Ke{\ss}ler}\ \emph {et~al.}(2021)\citenamefont {Ke{\ss}ler}, \citenamefont {Kongkhambut}, \citenamefont {Georges}, \citenamefont {Mathey}, \citenamefont {Cosme},\ and\ \citenamefont {Hemmerich}}]{9}%
  \BibitemOpen
  \bibfield  {author} {\bibinfo {author} {\bibfnamefont {H.}~\bibnamefont {Ke{\ss}ler}}, \bibinfo {author} {\bibfnamefont {P.}~\bibnamefont {Kongkhambut}}, \bibinfo {author} {\bibfnamefont {C.}~\bibnamefont {Georges}}, \bibinfo {author} {\bibfnamefont {L.}~\bibnamefont {Mathey}}, \bibinfo {author} {\bibfnamefont {J.~G.}\ \bibnamefont {Cosme}},\ and\ \bibinfo {author} {\bibfnamefont {A.}~\bibnamefont {Hemmerich}},\ }\bibfield  {title} {\bibinfo {title} {Observation of a dissipative time crystal},\ }\href@noop {} {\bibfield  {journal} {\bibinfo  {journal} {Phys. Rev. Lett.}\ }\textbf {\bibinfo {volume} {127}},\ \bibinfo {pages} {043602} (\bibinfo {year} {2021})}\BibitemShut {NoStop}%
\bibitem [{\citenamefont {Zhang}\ \emph {et~al.}(2017)\citenamefont {Zhang}, \citenamefont {Hess}, \citenamefont {Kyprianidis}, \citenamefont {Becker}, \citenamefont {Lee}, \citenamefont {Smith}, \citenamefont {Pagano}, \citenamefont {Potirniche}, \citenamefont {Potter}, \citenamefont {Vishwanath} \emph {et~al.}}]{10}%
  \BibitemOpen
  \bibfield  {author} {\bibinfo {author} {\bibfnamefont {J.}~\bibnamefont {Zhang}}, \bibinfo {author} {\bibfnamefont {P.~W.}\ \bibnamefont {Hess}}, \bibinfo {author} {\bibfnamefont {A.}~\bibnamefont {Kyprianidis}}, \bibinfo {author} {\bibfnamefont {P.}~\bibnamefont {Becker}}, \bibinfo {author} {\bibfnamefont {A.}~\bibnamefont {Lee}}, \bibinfo {author} {\bibfnamefont {J.}~\bibnamefont {Smith}}, \bibinfo {author} {\bibfnamefont {G.}~\bibnamefont {Pagano}}, \bibinfo {author} {\bibfnamefont {I.-D.}\ \bibnamefont {Potirniche}}, \bibinfo {author} {\bibfnamefont {A.~C.}\ \bibnamefont {Potter}}, \bibinfo {author} {\bibfnamefont {A.}~\bibnamefont {Vishwanath}}, \emph {et~al.},\ }\bibfield  {title} {\bibinfo {title} {Observation of a discrete time crystal},\ }\href@noop {} {\bibfield  {journal} {\bibinfo  {journal} {Nature}\ }\textbf {\bibinfo {volume} {543}},\ \bibinfo {pages} {217} (\bibinfo {year} {2017})}\BibitemShut {NoStop}%
\bibitem [{\citenamefont {Chen}\ \emph {et~al.}(2024)\citenamefont {Chen}, \citenamefont {Peng}, \citenamefont {Li}, \citenamefont {Chesi},\ and\ \citenamefont {Wang}}]{12}%
  \BibitemOpen
  \bibfield  {author} {\bibinfo {author} {\bibfnamefont {D.}~\bibnamefont {Chen}}, \bibinfo {author} {\bibfnamefont {Z.}~\bibnamefont {Peng}}, \bibinfo {author} {\bibfnamefont {J.}~\bibnamefont {Li}}, \bibinfo {author} {\bibfnamefont {S.}~\bibnamefont {Chesi}},\ and\ \bibinfo {author} {\bibfnamefont {Y.}~\bibnamefont {Wang}},\ }\bibfield  {title} {\bibinfo {title} {Discrete time crystal in an open optomechanical system},\ }\href@noop {} {\bibfield  {journal} {\bibinfo  {journal} {Phys. Rev. Res.}\ }\textbf {\bibinfo {volume} {6}},\ \bibinfo {pages} {013130} (\bibinfo {year} {2024})}\BibitemShut {NoStop}%
\bibitem [{\citenamefont {Rovny}\ \emph {et~al.}(2018)\citenamefont {Rovny}, \citenamefont {Blum},\ and\ \citenamefont {Barrett}}]{13}%
  \BibitemOpen
  \bibfield  {author} {\bibinfo {author} {\bibfnamefont {J.}~\bibnamefont {Rovny}}, \bibinfo {author} {\bibfnamefont {R.~L.}\ \bibnamefont {Blum}},\ and\ \bibinfo {author} {\bibfnamefont {S.~E.}\ \bibnamefont {Barrett}},\ }\bibfield  {title} {\bibinfo {title} {Observation of discrete-time-crystal signatures in an ordered dipolar many-body system},\ }\href@noop {} {\bibfield  {journal} {\bibinfo  {journal} {Phys. Rev. Lett.}\ }\textbf {\bibinfo {volume} {120}},\ \bibinfo {pages} {180603} (\bibinfo {year} {2018})}\BibitemShut {NoStop}%
\bibitem [{\citenamefont {Pal}\ \emph {et~al.}(2018)\citenamefont {Pal}, \citenamefont {Nishad}, \citenamefont {Mahesh},\ and\ \citenamefont {Sreejith}}]{14}%
  \BibitemOpen
  \bibfield  {author} {\bibinfo {author} {\bibfnamefont {S.}~\bibnamefont {Pal}}, \bibinfo {author} {\bibfnamefont {N.}~\bibnamefont {Nishad}}, \bibinfo {author} {\bibfnamefont {T.}~\bibnamefont {Mahesh}},\ and\ \bibinfo {author} {\bibfnamefont {G.}~\bibnamefont {Sreejith}},\ }\bibfield  {title} {\bibinfo {title} {Temporal order in periodically driven spins in star-shaped clusters},\ }\href@noop {} {\bibfield  {journal} {\bibinfo  {journal} {Phys. Rev. Lett.}\ }\textbf {\bibinfo {volume} {120}},\ \bibinfo {pages} {180602} (\bibinfo {year} {2018})}\BibitemShut {NoStop}%
\bibitem [{\citenamefont {Tucker}\ \emph {et~al.}(2018)\citenamefont {Tucker}, \citenamefont {Zhu}, \citenamefont {Lewis-Swan}, \citenamefont {Marino}, \citenamefont {Jimenez}, \citenamefont {Restrepo},\ and\ \citenamefont {Rey}}]{36}%
  \BibitemOpen
  \bibfield  {author} {\bibinfo {author} {\bibfnamefont {K.}~\bibnamefont {Tucker}}, \bibinfo {author} {\bibfnamefont {B.}~\bibnamefont {Zhu}}, \bibinfo {author} {\bibfnamefont {R.~J.}\ \bibnamefont {Lewis-Swan}}, \bibinfo {author} {\bibfnamefont {J.}~\bibnamefont {Marino}}, \bibinfo {author} {\bibfnamefont {F.}~\bibnamefont {Jimenez}}, \bibinfo {author} {\bibfnamefont {J.~G.}\ \bibnamefont {Restrepo}},\ and\ \bibinfo {author} {\bibfnamefont {A.~M.}\ \bibnamefont {Rey}},\ }\bibfield  {title} {\bibinfo {title} {Shattered time: can a dissipative time crystal survive many-body correlations?},\ }\href@noop {} {\bibfield  {journal} {\bibinfo  {journal} {New J. Phys.}\ }\textbf {\bibinfo {volume} {20}},\ \bibinfo {pages} {123003} (\bibinfo {year} {2018})}\BibitemShut {NoStop}%
\bibitem [{\citenamefont {Bu{\v{c}}a}\ \emph {et~al.}(2019)\citenamefont {Bu{\v{c}}a}, \citenamefont {Tindall},\ and\ \citenamefont {Jaksch}}]{37}%
  \BibitemOpen
  \bibfield  {author} {\bibinfo {author} {\bibfnamefont {B.}~\bibnamefont {Bu{\v{c}}a}}, \bibinfo {author} {\bibfnamefont {J.}~\bibnamefont {Tindall}},\ and\ \bibinfo {author} {\bibfnamefont {D.}~\bibnamefont {Jaksch}},\ }\bibfield  {title} {\bibinfo {title} {Non-stationary coherent quantum many-body dynamics through dissipation},\ }\href@noop {} {\bibfield  {journal} {\bibinfo  {journal} {Nat. Commun.}\ }\textbf {\bibinfo {volume} {10}},\ \bibinfo {pages} {1730} (\bibinfo {year} {2019})}\BibitemShut {NoStop}%
\bibitem [{\citenamefont {Booker}\ \emph {et~al.}(2020)\citenamefont {Booker}, \citenamefont {Bu{\v{c}}a},\ and\ \citenamefont {Jaksch}}]{27}%
  \BibitemOpen
  \bibfield  {author} {\bibinfo {author} {\bibfnamefont {C.}~\bibnamefont {Booker}}, \bibinfo {author} {\bibfnamefont {B.}~\bibnamefont {Bu{\v{c}}a}},\ and\ \bibinfo {author} {\bibfnamefont {D.}~\bibnamefont {Jaksch}},\ }\bibfield  {title} {\bibinfo {title} {Non-stationarity and dissipative time crystals: spectral properties and finite-size effects},\ }\href@noop {} {\bibfield  {journal} {\bibinfo  {journal} {New J. Phys.}\ }\textbf {\bibinfo {volume} {22}},\ \bibinfo {pages} {085007} (\bibinfo {year} {2020})}\BibitemShut {NoStop}%
\bibitem [{\citenamefont {Seibold}\ \emph {et~al.}(2020)\citenamefont {Seibold}, \citenamefont {Rota},\ and\ \citenamefont {Savona}}]{38}%
  \BibitemOpen
  \bibfield  {author} {\bibinfo {author} {\bibfnamefont {K.}~\bibnamefont {Seibold}}, \bibinfo {author} {\bibfnamefont {R.}~\bibnamefont {Rota}},\ and\ \bibinfo {author} {\bibfnamefont {V.}~\bibnamefont {Savona}},\ }\bibfield  {title} {\bibinfo {title} {Dissipative time crystal in an asymmetric nonlinear photonic dimer},\ }\href@noop {} {\bibfield  {journal} {\bibinfo  {journal} {Phys. Rev. A}\ }\textbf {\bibinfo {volume} {101}},\ \bibinfo {pages} {033839} (\bibinfo {year} {2020})}\BibitemShut {NoStop}%
\bibitem [{\citenamefont {Taheri}\ \emph {et~al.}(2022)\citenamefont {Taheri}, \citenamefont {Matsko}, \citenamefont {Maleki},\ and\ \citenamefont {Sacha}}]{39}%
  \BibitemOpen
  \bibfield  {author} {\bibinfo {author} {\bibfnamefont {H.}~\bibnamefont {Taheri}}, \bibinfo {author} {\bibfnamefont {A.~B.}\ \bibnamefont {Matsko}}, \bibinfo {author} {\bibfnamefont {L.}~\bibnamefont {Maleki}},\ and\ \bibinfo {author} {\bibfnamefont {K.}~\bibnamefont {Sacha}},\ }\bibfield  {title} {\bibinfo {title} {All-optical dissipative discrete time crystals},\ }\href@noop {} {\bibfield  {journal} {\bibinfo  {journal} {Nat. Commun.}\ }\textbf {\bibinfo {volume} {13}},\ \bibinfo {pages} {848} (\bibinfo {year} {2022})}\BibitemShut {NoStop}%
\bibitem [{\citenamefont {Vu}\ and\ \citenamefont {Das~Sarma}(2023)}]{40}%
  \BibitemOpen
  \bibfield  {author} {\bibinfo {author} {\bibfnamefont {D.}~\bibnamefont {Vu}}\ and\ \bibinfo {author} {\bibfnamefont {S.}~\bibnamefont {Das~Sarma}},\ }\bibfield  {title} {\bibinfo {title} {Dissipative prethermal discrete time crystal},\ }\href@noop {} {\bibfield  {journal} {\bibinfo  {journal} {Phys. Rev. Lett.}\ }\textbf {\bibinfo {volume} {130}},\ \bibinfo {pages} {130401} (\bibinfo {year} {2023})}\BibitemShut {NoStop}%
\bibitem [{\citenamefont {Iemini}\ \emph {et~al.}(2018)\citenamefont {Iemini}, \citenamefont {Russomanno}, \citenamefont {Keeling}, \citenamefont {Schir{\`o}}, \citenamefont {Dalmonte},\ and\ \citenamefont {Fazio}}]{41}%
  \BibitemOpen
  \bibfield  {author} {\bibinfo {author} {\bibfnamefont {F.}~\bibnamefont {Iemini}}, \bibinfo {author} {\bibfnamefont {A.}~\bibnamefont {Russomanno}}, \bibinfo {author} {\bibfnamefont {J.}~\bibnamefont {Keeling}}, \bibinfo {author} {\bibfnamefont {M.}~\bibnamefont {Schir{\`o}}}, \bibinfo {author} {\bibfnamefont {M.}~\bibnamefont {Dalmonte}},\ and\ \bibinfo {author} {\bibfnamefont {R.}~\bibnamefont {Fazio}},\ }\bibfield  {title} {\bibinfo {title} {Boundary time crystals},\ }\href@noop {} {\bibfield  {journal} {\bibinfo  {journal} {Phys. Rev. Lett.}\ }\textbf {\bibinfo {volume} {121}},\ \bibinfo {pages} {035301} (\bibinfo {year} {2018})}\BibitemShut {NoStop}%
\bibitem [{\citenamefont {Carollo}\ \emph {et~al.}(2020)\citenamefont {Carollo}, \citenamefont {Brandner},\ and\ \citenamefont {Lesanovsky}}]{42}%
  \BibitemOpen
  \bibfield  {author} {\bibinfo {author} {\bibfnamefont {F.}~\bibnamefont {Carollo}}, \bibinfo {author} {\bibfnamefont {K.}~\bibnamefont {Brandner}},\ and\ \bibinfo {author} {\bibfnamefont {I.}~\bibnamefont {Lesanovsky}},\ }\bibfield  {title} {\bibinfo {title} {Nonequilibrium many-body quantum engine driven by time-translation symmetry breaking},\ }\href@noop {} {\bibfield  {journal} {\bibinfo  {journal} {Phys. Rev. Lett.}\ }\textbf {\bibinfo {volume} {125}},\ \bibinfo {pages} {240602} (\bibinfo {year} {2020})}\BibitemShut {NoStop}%
\bibitem [{\citenamefont {Prazeres}\ \emph {et~al.}(2021)\citenamefont {Prazeres}, \citenamefont {Souza},\ and\ \citenamefont {Iemini}}]{43}%
  \BibitemOpen
  \bibfield  {author} {\bibinfo {author} {\bibfnamefont {L.~F.~d.}\ \bibnamefont {Prazeres}}, \bibinfo {author} {\bibfnamefont {L.~d.~S.}\ \bibnamefont {Souza}},\ and\ \bibinfo {author} {\bibfnamefont {F.}~\bibnamefont {Iemini}},\ }\bibfield  {title} {\bibinfo {title} {Boundary time crystals in collective d-level systems},\ }\href@noop {} {\bibfield  {journal} {\bibinfo  {journal} {Phys. Rev. B}\ }\textbf {\bibinfo {volume} {103}},\ \bibinfo {pages} {184308} (\bibinfo {year} {2021})}\BibitemShut {NoStop}%
\bibitem [{\citenamefont {Piccitto}\ \emph {et~al.}(2021)\citenamefont {Piccitto}, \citenamefont {Wauters}, \citenamefont {Nori},\ and\ \citenamefont {Shammah}}]{44}%
  \BibitemOpen
  \bibfield  {author} {\bibinfo {author} {\bibfnamefont {G.}~\bibnamefont {Piccitto}}, \bibinfo {author} {\bibfnamefont {M.}~\bibnamefont {Wauters}}, \bibinfo {author} {\bibfnamefont {F.}~\bibnamefont {Nori}},\ and\ \bibinfo {author} {\bibfnamefont {N.}~\bibnamefont {Shammah}},\ }\bibfield  {title} {\bibinfo {title} {Symmetries and conserved quantities of boundary time crystals in generalized spin models},\ }\href@noop {} {\bibfield  {journal} {\bibinfo  {journal} {Phys. Rev. B}\ }\textbf {\bibinfo {volume} {104}},\ \bibinfo {pages} {014307} (\bibinfo {year} {2021})}\BibitemShut {NoStop}%
\bibitem [{\citenamefont {Carollo}\ and\ \citenamefont {Lesanovsky}(2022)}]{45}%
  \BibitemOpen
  \bibfield  {author} {\bibinfo {author} {\bibfnamefont {F.}~\bibnamefont {Carollo}}\ and\ \bibinfo {author} {\bibfnamefont {I.}~\bibnamefont {Lesanovsky}},\ }\bibfield  {title} {\bibinfo {title} {Exact solution of a boundary time-crystal phase transition: time-translation symmetry breaking and non-markovian dynamics of correlations},\ }\href@noop {} {\bibfield  {journal} {\bibinfo  {journal} {Phys. Rev. A}\ }\textbf {\bibinfo {volume} {105}},\ \bibinfo {pages} {L040202} (\bibinfo {year} {2022})}\BibitemShut {NoStop}%
\bibitem [{\citenamefont {Sergeev}\ \emph {et~al.}(2024)\citenamefont {Sergeev}, \citenamefont {Zyablovsky}, \citenamefont {Andrianov},\ and\ \citenamefont {Lozovik}}]{15}%
  \BibitemOpen
  \bibfield  {author} {\bibinfo {author} {\bibfnamefont {T.~T.}\ \bibnamefont {Sergeev}}, \bibinfo {author} {\bibfnamefont {A.~A.}\ \bibnamefont {Zyablovsky}}, \bibinfo {author} {\bibfnamefont {E.~S.}\ \bibnamefont {Andrianov}},\ and\ \bibinfo {author} {\bibfnamefont {Y.~E.}\ \bibnamefont {Lozovik}},\ }\bibfield  {title} {\bibinfo {title} {Spontaneous breaking of time translation symmetry in a system without periodic external driving},\ }\href@noop {} {\bibfield  {journal} {\bibinfo  {journal} {Opt. Lett.}\ }\textbf {\bibinfo {volume} {49}},\ \bibinfo {pages} {4783} (\bibinfo {year} {2024})}\BibitemShut {NoStop}%
\bibitem [{\citenamefont {Scully}\ and\ \citenamefont {Zubairy}(1999)}]{23}%
  \BibitemOpen
  \bibfield  {author} {\bibinfo {author} {\bibfnamefont {M.~O.}\ \bibnamefont {Scully}}\ and\ \bibinfo {author} {\bibfnamefont {M.~S.}\ \bibnamefont {Zubairy}},\ }\href@noop {} {\emph {\bibinfo {title} {Quantum optics}}}\ (\bibinfo  {publisher} {Cambridge University Press},\ \bibinfo {year} {1999})\BibitemShut {NoStop}%
\bibitem [{\citenamefont {Berestetskii}\ \emph {et~al.}(1982)\citenamefont {Berestetskii}, \citenamefont {Lifshitz},\ and\ \citenamefont {Pitaevskii}}]{LL4}%
  \BibitemOpen
  \bibfield  {author} {\bibinfo {author} {\bibfnamefont {V.~B.}\ \bibnamefont {Berestetskii}}, \bibinfo {author} {\bibfnamefont {E.~M.}\ \bibnamefont {Lifshitz}},\ and\ \bibinfo {author} {\bibfnamefont {L.~P.}\ \bibnamefont {Pitaevskii}},\ }\href@noop {} {\emph {\bibinfo {title} {Quantum Electrodynamics: Volume 4}}},\ Vol.~\bibinfo {volume} {4}\ (\bibinfo  {publisher} {Butterworth-Heinemann},\ \bibinfo {year} {1982})\BibitemShut {NoStop}%
\bibitem [{\citenamefont {T{\'o}th}\ and\ \citenamefont {Petz}(2013)}]{toth2013}%
  \BibitemOpen
  \bibfield  {author} {\bibinfo {author} {\bibfnamefont {G.}~\bibnamefont {T{\'o}th}}\ and\ \bibinfo {author} {\bibfnamefont {D.}~\bibnamefont {Petz}},\ }\bibfield  {title} {\bibinfo {title} {Extremal properties of the variance and the quantum fisher information},\ }\href@noop {} {\bibfield  {journal} {\bibinfo  {journal} {Physical Review A—Atomic, Molecular, and Optical Physics}\ }\textbf {\bibinfo {volume} {87}},\ \bibinfo {pages} {032324} (\bibinfo {year} {2013})}\BibitemShut {NoStop}%
\bibitem [{\citenamefont {Wang}\ \emph {et~al.}(2014)\citenamefont {Wang}, \citenamefont {Wu}, \citenamefont {Yang}, \citenamefont {Jin}, \citenamefont {Lambert},\ and\ \citenamefont {Nori}}]{wang2014}%
  \BibitemOpen
  \bibfield  {author} {\bibinfo {author} {\bibfnamefont {T.-L.}\ \bibnamefont {Wang}}, \bibinfo {author} {\bibfnamefont {L.-N.}\ \bibnamefont {Wu}}, \bibinfo {author} {\bibfnamefont {W.}~\bibnamefont {Yang}}, \bibinfo {author} {\bibfnamefont {G.-R.}\ \bibnamefont {Jin}}, \bibinfo {author} {\bibfnamefont {N.}~\bibnamefont {Lambert}},\ and\ \bibinfo {author} {\bibfnamefont {F.}~\bibnamefont {Nori}},\ }\bibfield  {title} {\bibinfo {title} {Quantum fisher information as a signature of the superradiant quantum phase transition},\ }\href@noop {} {\bibfield  {journal} {\bibinfo  {journal} {New J. Phys.}\ }\textbf {\bibinfo {volume} {16}},\ \bibinfo {pages} {063039} (\bibinfo {year} {2014})}\BibitemShut {NoStop}%
\bibitem [{\citenamefont {Liu}\ \emph {et~al.}(2013)\citenamefont {Liu}, \citenamefont {Ma},\ and\ \citenamefont {Wang}}]{liu2013}%
  \BibitemOpen
  \bibfield  {author} {\bibinfo {author} {\bibfnamefont {W.-F.}\ \bibnamefont {Liu}}, \bibinfo {author} {\bibfnamefont {J.}~\bibnamefont {Ma}},\ and\ \bibinfo {author} {\bibfnamefont {X.}~\bibnamefont {Wang}},\ }\bibfield  {title} {\bibinfo {title} {Quantum fisher information and spin squeezing in the ground state of the xy model},\ }\href@noop {} {\bibfield  {journal} {\bibinfo  {journal} {J. Phys. A-Math.}\ }\textbf {\bibinfo {volume} {46}},\ \bibinfo {pages} {045302} (\bibinfo {year} {2013})}\BibitemShut {NoStop}%
\bibitem [{\citenamefont {Ma}\ and\ \citenamefont {Wang}(2009)}]{ma2009}%
  \BibitemOpen
  \bibfield  {author} {\bibinfo {author} {\bibfnamefont {J.}~\bibnamefont {Ma}}\ and\ \bibinfo {author} {\bibfnamefont {X.}~\bibnamefont {Wang}},\ }\bibfield  {title} {\bibinfo {title} {Fisher information and spin squeezing in the lipkin-meshkov-glick model},\ }\href@noop {} {\bibfield  {journal} {\bibinfo  {journal} {Phys. Rev. A}\ }\textbf {\bibinfo {volume} {80}},\ \bibinfo {pages} {012318} (\bibinfo {year} {2009})}\BibitemShut {NoStop}%
\end{thebibliography}%

\end{document}